# The angular spectrum representation of vectorial laser beams

**Peter Muys**

*Laser Power Optics*
*Hooiland 3, B-9030 Gent Belgium*
*peter.muys@gmail.com*

The angular spectrum of a vectorial laser beam is expressed in terms of an intrinsic coordinate system instead of the usual Cartesian laboratory coordinates. This switch leads to simple, elegant and new expressions such as for the angular spectrum of the Hertz vector corresponding to the electromagnetic fields. As an application of this approach, we consider axially symmetric vector beams, showing non-diffracting properties of these beams, without invoking the paraxial approximation. Further, we indicate the relevance of the method for analyzing nonparaxial resonators, such as microresonators.

The angular spectrum representation of a vectorial laser field as solution of the vectorial Helmholtz equation is now a well known topic [1,2]. The drawback however is that the resulting formulas are complicated, due to the appearance of the Cartesian base vectors in them. In this paper, we lift this constraint, by changing to new coordinates, which renders the formulas shorter and more transparent to physical interpretation. In a second stage, we implement the method to the angular spectrum of the Hertz vector. This leads to an alternative angular spectrum representation of the Helmholtz equation. As an application of the usefulness of this representation, we will briefly discuss diffraction-free vectorial beams.

Let **k** be the wave vector of a plane wave

$$\mathbf{k} = k_x \mathbf{e}_x + k_y \mathbf{e}_y + k_z \mathbf{e}_z$$
$$= k_t \mathbf{e}_r + k_z \mathbf{e}_z \quad (1)$$
$$= k_t \cos\varphi \, \mathbf{e}_x + k_t \sin\varphi \, \mathbf{e}_y + k_z \mathbf{e}_z$$

and its magnitude is given by

$$k^2 = k_x^2 + k_y^2 + k_z^2 = k_t^2 + k_z^2$$

In eq.(1) we have denoted the split of the wave vector into its transverse (subscript t) and longitudinal part (subscript z). The k-vector forms together with the z-axis the meridional plane [3]. In optical terminology, we would call it the plane of incidence (on a mirror for example, where z is the normal to the mirror). The angle between both axes is the polar angle θ. The cylindrical coordinates are defined through their unit vectors $\mathbf{e}_r$, $\mathbf{e}_z$, and $\mathbf{e}_\varphi$, φ being the azimuthal angle of the k-vector in the transverse (x,y) plane.

The three-dimensional locus vector **R** is defined by

$$\mathbf{R} = x\mathbf{e}_x + y\mathbf{e}_y + z\mathbf{e}_z$$
$$= r\mathbf{e}_r + z\mathbf{e}_z \quad (2)$$
$$= r\cos\alpha \, \mathbf{e}_x + r\sin\alpha \, \mathbf{e}_y + z\mathbf{e}_z$$

where we indicated the split of the locus vector in its transverse and its longitudinal part.

The key element in our approach is the point brought up in ref. [2] and [4], that we should consider an *intrinsic* coordinate system. It contains three unit vectors, two of them are resp. perpendicular and parallel to the meridional plane, the third being the unit vector in the k direction:

$$\mathbf{s} = \mathbf{e}_\varphi$$
$$\mathbf{p} = \frac{k_z}{k}\mathbf{e}_r - \frac{k_t}{k}\mathbf{e}_z \quad (3)$$
$$\mathbf{m} = \mathbf{k}/k$$

These unit vectors, taken in this order, form a right-hand sided coordinate system.

Note that **s** does not possess a z-component. **s** is perpendicular to the meridional plane, whereas **p** is coplanar with it, just like **m**. **s** has only a transverse component, its azimuthal part, whereas **p** both has a transverse component, its radial part, and a longitudinal. An alternative but equivalent definition of the intrinsic coordinates is:

$$\mathbf{s} = \frac{\mathbf{e}_z \times \mathbf{k}}{|\mathbf{e}_z \times \mathbf{k}|} = \frac{\mathbf{e}_z \times \mathbf{k}}{k_t}$$

$$\mathbf{p} = \frac{\mathbf{s} \times \mathbf{k}}{k} = \mathbf{s} \times \mathbf{m}$$

The inverse transformation from the **s,p,m** intrinsic coordinate system to the Cartesian laboratory coordinate system $\mathbf{e}_x$, $\mathbf{e}_y$, $\mathbf{e}_z$ is, after lengthy but straightforward calculations, given by

$$\mathbf{e}_x = -\frac{k_y}{k_t}\mathbf{s} + \frac{k_x k_z}{k\, k_t}\mathbf{p} + \frac{k_x}{k}\mathbf{m}$$

$$\mathbf{e}_y = \frac{k_x}{k_t}\mathbf{s} + \frac{k_y k_z}{k\, k_t}\mathbf{p} + \frac{k_y}{k}\mathbf{m} \quad (4)$$

$$\mathbf{e}_z = \qquad\qquad -\frac{k_t}{k}\mathbf{p} + \frac{k_z}{k}\mathbf{m}$$

which is a central result for the subsequent calculations.

We now turn to introduce the Hertz vector potential. Here, the starting assumption is that the (electric) Hertz vector **P** takes on a scalar form:

$$P_z = \frac{1}{k^2} V(x,y,z,t)$$
$$P_x = P_y = 0 \quad (5)$$

This supposition regarding the scalar character of the Hertz vector potential is identical to the one adopted in [3]. The attractive aspect of the Hertz vector representation as compared to the modal representation is that it arrives at the same rigorous results as the modal representation, but in a much shorter and more elegant way, because, although sounding at first contradicting, the Hertz *vector* in many practical cases reduces to a *scalar*. To give a quick example, the electric field of a radiating electric dipole is dependent on the considered radiation direction, but the Hertz vector is nevertheless everywhere in space a scalar, independent of the radiation direction [3].

In eq.(5), we now assume the polarization of the Hertz vector to be along the z-axis, as in ref.[5] . This is opposed to ref [6] where the polarization was taken along the x axis.

V(x,y,z,t) is a scalar solution of the wave equation with the boundary condition V(x,y,0,t)=v(x,y) exp(-jωt). The Fourier transform of the Hertz potential v(x,y) is given by

$$F(k_x,k_y) = \frac{1}{2\pi} \int_{-\infty}^{\infty}\int_{-\infty}^{\infty} v(x,y)\exp\left[-j(k_x x + k_y y)\right] dx\, dy \quad (6)$$

and is called its angular spectrum (in the plane z=0). Note that F is *not* the angular spectrum of the electric field in the plane z=0, but of the Hertz vector. The general solution of the wave equation in the half-space z>0 is now [7]

$$V(x,y,z,t) = \frac{\exp(-j\omega t)}{2\pi} \times \int_{-\infty}^{\infty}\int_{-\infty}^{\infty} F(k_x,k_y)\exp\left[j(k_x x + k_y y + k_z z)\right] dk_x dk_y \quad (7)$$

which we can substitute in the vectorial equation linking the electrical field to the Hertz vector [6]:

$$\mathbf{E}(\mathbf{R}) = -\frac{\varepsilon\mu}{c^2}\frac{\partial^2 \mathbf{P}}{\partial t^2} + \text{grad div } \mathbf{P}$$

The angular spectrum representation of the electric field becomes

$$\mathbf{E}(\mathbf{R}) = \frac{\exp(-j\omega t)}{2\pi k^2} \times \int_{-\infty}^{\infty}\int_{-\infty}^{\infty} \mathbf{g}(k_x,k_y)\, F(k_x,k_y)\exp(j\mathbf{k}\cdot\mathbf{R}) dk_x dk_y \quad (8)$$

For the magnetic field, a similar procedure [6] delivers

$$\mathbf{B}(\mathbf{R}) = \frac{\varepsilon\mu}{c}\text{rot}(\frac{\partial \mathbf{P}}{\partial t}) = \frac{\exp(-j\omega t)}{2\pi k}\sqrt{\varepsilon\mu} \times \int_{-\infty}^{\infty}\int_{-\infty}^{\infty} \mathbf{h}(k_x,k_y)\, F(k_x,k_y)\exp(j\mathbf{k}\cdot\mathbf{R})\, dk_x dk_y \quad (9)$$

The vectorial parts of the integrals in eqs.(8) and (9) come out to be

$$\mathbf{g}(k_x,k_y) = (k^2 - k_z^2)\, \mathbf{e}_z - k_x k_z \mathbf{e}_x - k_y k_z \mathbf{e}_y$$
$$\mathbf{h}(k_x,k_y) = k_y \mathbf{e}_x - k_x \mathbf{e}_y$$

Expressed in intrinsic coordinates, the fields given by eqs. (8) and (9) become by using eq. (4), and by suppressing the harmonic time dependency

$$\mathbf{E}(\mathbf{R}) = \frac{-1}{2\pi}\int_{-\infty}^{\infty}\int_{-\infty}^{\infty} \frac{k_t}{k}\mathbf{p}\, F(k_x,k_y)\exp(j\mathbf{k}\cdot\mathbf{R})\, dk_x dk_y \quad (10)$$

$$\mathbf{B}(\mathbf{R}) = \frac{-1}{2\pi}\sqrt{\varepsilon\mu}\int_{-\infty}^{\infty}\int_{-\infty}^{\infty} \frac{k_t}{k}\mathbf{s}\, F(k_x,k_y)\exp(j\mathbf{k}\cdot\mathbf{R})\, dk_x dk_y$$

The magnetic field vector only depends on **s**, which has no z-component, see eq.(3), so hence the fields (10) are TM. For the corresponding TE fields, we need to consider the magnetic Hertz vector oriented parallel to the z-axis.

Following another strategy to solve the vectorial wave equation, ref.[1] started from the modal representation, as commonly used in guided wave problems, combining it with the angular spectrum representation. The advantage of this approach is that it clearly identifies and separates the contributions of the transverse fields (TE and TM) right from the start. A similar procedure of transforming to intrinsic coordinates can be executed on the formulas of ref.[1] , which are originally presented in laboratory Cartesian coordinates as

$$\mathbf{E}_{TM}(\mathbf{R}) = \frac{-1}{4\pi^2} \times$$
$$\int_{-\infty}^{\infty}\int_{-\infty}^{\infty}\left[\frac{k_z}{kk_t}(k_x \mathbf{e}_x + k_y \mathbf{e}_y) - \frac{k_t}{k}\mathbf{e}_z\right] A_{TM}(k_x,k_y)\exp(-j\mathbf{k}\cdot\mathbf{R}) dk_x dk_y$$

$$\mathbf{E}_{TE}(\mathbf{R}) = \frac{-1}{4\pi^2} \times$$
$$\int_{-\infty}^{\infty}\int_{-\infty}^{\infty}\left[\frac{k_y}{k_t}\mathbf{e}_x - \frac{k_x}{k_t}\mathbf{e}_y\right] A_{TE}(k_x,k_y)\exp(-j\mathbf{k}\cdot\mathbf{R}) dk_x dk_y$$

So ref.[1] was able to separate the propagating field into two transverse contributions, which ref.[2,4] did not indicate in this way. We will see however, that in ref.[4] essentially the same result was obtained. Expressions for the magnetic field are not given by ref.[1]. It is nevertheless clear that $\mathbf{E}_{TE}$ is lacking a z-component, as it should for the electric field of a TE mode. Again using the transformations (4), these expressions now take on a very simple form:

$$\mathbf{E}_{TM}(\mathbf{R}) = \frac{-1}{4\pi^2}\int_{-\infty}^{\infty}\int_{-\infty}^{\infty} \mathbf{p}\, A_{TM}(k_x,k_y)\exp(-j\mathbf{k}\cdot\mathbf{R}) dk_x dk_y$$
$$\mathbf{E}_{TE}(\mathbf{R}) = \frac{-1}{4\pi^2}\int_{-\infty}^{\infty}\int_{-\infty}^{\infty} \mathbf{s}\, A_{TE}(k_x,k_y)\exp(-j\mathbf{k}\cdot\mathbf{R}) dk_x dk_y \quad (11)$$

so that the vectorial angular spectrum of the total vectorial electric field $\mathbf{E}(\mathbf{R}) = \mathbf{E}_{TM}(\mathbf{R}) + \mathbf{E}_{TE}(\mathbf{R})$
is given by

$$\mathbf{A}(k_x,k_y) = A_{TM}\,\mathbf{p} + A_{TE}\,\mathbf{s} \quad (12)$$

which is exactly the form given in ref.[4], although in a slightly different format, denoted as $\mathbf{A} = A_p\,\mathbf{p} + A_s\,\mathbf{s}$ without explicitly pointing to the transverse character of the contributions $A_s$ and $A_p$.

Just as in eq.(6), the spectra $A_{TM}$ and $A_{TE}$ are in fact 2D Fourier Transforms defined by the fields in z=0 [1].

Eqs.(10) and (11) are, besides for their compactness, remarkable in a further way. The denomination "TE" and "TM" indicate that the fields are transverse to the *fixed* z-axis. Notwithstanding, the TE electric field can be expressed as a function of the *variable* s-vector, and the TM electric field as a function of the *variable* p-vector.

By comparing eqs. (10) and (11), we can link the angular spectrum of the electrical Hertz vector to the angular spectrum of the TM field:

$$A_{TM}(k_x,k_y) = \frac{2\pi k_t}{k} F(k_x,k_y) \qquad (13)$$

We now work out the general eqs.(11) for the special case of axial symmetry of the fields. Both ref. [1] and [2] have considered this case. The angular spectrum is independent of the azimuthal angle $\varphi$ in reciprocal k-space and only depends on $k_t$. Eqs.(11) become [1]:

$$\mathbf{E}_{TM}(\mathbf{R}) = \frac{1}{2\pi k} \times$$
$$\int_0^k \left[ jk_z J_1(k_t r)\, \mathbf{e}_r - k_t J_0(k_t r)\, \mathbf{e}_z \right] k_t A_{TM}(k_t) \exp[-jk_z z]\, dk_t$$

$$\mathbf{E}_{TE}(\mathbf{R}) = \frac{-j\mathbf{e}_\alpha}{2\pi} \times$$
$$\int_0^k A_{TE}(k_t) J_1(k_t r)\, k_t \exp[-jk_z z]\, dk_t$$

(14)

where the angular spectra are given by Hankel transforms:

$$A_{TM}(k_t) = \frac{2\pi k j}{\sqrt{k^2 - k_t^2}} \int_0^\infty E_r(r,z=0) J_1(k_t r) r\, dr$$
$$A_{TE}(k_t) = -2\pi j \int_0^\infty E_\varphi(r,z=0) J_1(k_t r) r\, dr \qquad (15)$$

$J_0(z)$ and $J_1(z)$ are the well known Bessel functions. If we rewrite eqs.(12) and (14) by just keeping the vectorial part and the differentials in place, and absorbing the other contributions in the symbol (.), we can more clearly see how the unit vectors are transformed by changing from Cartesian to cylindrical coordinates. Note that $\mathbf{e}_\varphi$ as integration variable in k-space is transformed into $\mathbf{e}_\alpha$ in R-space, which is independent of $k_x$ and $k_y$ and hence can be brought in front of the integration sign:

$$\mathbf{E}_{TE} = \iint \mathbf{s}\, (.)dk_x dk_y = \iint \mathbf{e}_\varphi (.) dk_x dk_y = \mathbf{e}_\alpha \int (.) dk_t$$

$$\mathbf{E}_{TM} = \iint \mathbf{p}\, (.)dk_x dk_y = \iint (\frac{k_z}{k}\mathbf{e}_r - \frac{k_t}{k}\mathbf{e}_z)\, (.) dk_x dk_y$$
$$= \int (J_1 \frac{k_z}{k}\mathbf{e}_r - J_0 \frac{k_t}{k}\mathbf{e}_z)\, (.) dk_t$$

The TM-vector keeps its transverse and longitudinal contributions, although now with other weighting coefficients, which are Bessel functions.

As a short application of eqs.(14,15), we will consider the diffraction-free propagation of an azimuthally polarized beam.

The general solution of the scalar wave equation for a diffraction-free beam is given [8] as:

$$u(x,y,z) = \exp(-jk_z z) \int_0^{2\pi} a(\varphi) \exp[jk_t(x\cos\varphi + y\sin\varphi)]\, d\varphi$$

and is known under the name "Whittaker integral" in the mathematical literature. Physically, it represents a superposition of plane waves with amplitude coefficient $a(\varphi)$ and with their wave vector situated on a cone with base radius $k_t$ and height $k_z$. In other words, their angular spectrum is given by $A = \delta(k_t - k_t')$. In the vectorial case, we use this same angular spectrum to describe nondiffracting beams, and substitute it in the TE-field expression of eq.(14) to arrive at

$$\mathbf{E}_{TE}(r,z) = \frac{-j}{2\pi} J_1(k_t r) \exp(-jk_z z)\, \mathbf{e}_\alpha \qquad (16)$$

The vectorial Bessel beam (16) is hence nondiffracting since the transverse part is not depending on the propagation coordinate z. This result was also obtained in [9], but by the method of separation of the variables of the paraxial wave equation. A mathematically equivalent statement is that the $J_1$ Bessel function can be represented as a Whittaker integral. Its explicit form can be found in ref.[3].

In summary, we have pointed out the relevance of using an intrinsic coordinate system to unify the existing angular spectrum methods. We have given the transformation formulas from intrinsic to Cartesian coordinates. The introduction of intrinsic coordinates much simplifies the angular spectral representation of vectorial beams. We have introduced the Hertz vector potential and have combined it with the intrinsic coordinates to represent TE and TM beams. This leads also to the mathematical link between the angular spectra of the fields and of their Hertz vector. Next, we derived expressions for a diffraction-free vectorial beam, without invoking the paraxial approximation. Finally, these results can be of importance to analyse nonparaxial resonators, such as microresonators.